\documentclass[epj]{svjour}
\usepackage{latexsym}
\usepackage{graphics}
\usepackage{amsmath}
\usepackage{epsfig} 

\begin{document}

\title{Temperature Dependence of Gluon and Ghost Propagators 
in Landau-Gauge Yang--Mills Theory below the Phase Transition}
\author{Burghard Gr{\"u}ter \inst{1} \and Reinhard Alkofer\inst{1} 
\and Axel Maas\inst{2}  \and Jochen Wambach\inst{2,3}}
%\thanks is optional - remove next line if not needed
%\thanks{\emph{Present address:} Insert the address here if needed}%
% Do not remove
%
%\offprints{Burghard Gr{\"u}ter}          % Insert a name or remove this line
%
\institute{Inst. f{\"u}r Theoretische Physik, Universit{\"a}t T{\"u}bingen, 
Auf der Morgenstelle 14, 72076 T{\"u}bingen, Germany \and 
Inst. f{\"u}r Kernphysik, Technische Universit{\"a}t Darmstadt, 
Schlo{\ss}gartenstr. 9, 64289 Darmstadt, Germany \and 
Gesellschaft f\"ur Schwerionenforschung mbH, Planckstr. 1, 
D-64291 Darmstadt, Germany} 
\date{\today }
% The correct dates will be entered by Springer
%
\abstract{
The Dyson--Schwinger equations of Landau-gauge Yang--Mills theory for the gluon
and ghost propagators are investigated. Numerical results are obtained within a
truncation scheme which has proven to be successful at vanishing temperature.
For temperatures up to 250 MeV we find only minor quantitative changes in the
infrared behaviour of the gluon and ghost propagators. The effective action
calculated from these propagators is temperature-independent within the 
numerical uncertainty.
 % end of PACS codes
} %end of abstract
\maketitle
\markboth{Burghard Gr{\"u}ter et al.: Temperature Dependence of Gluon and Ghost Propagators}
{Burghard Gr{\"u}ter et al.: Temperature Dependence of Gluon and Ghost Propagators}

\section{Introduction}
\label{sec:intro}

It is by now well established that for increasing temperature QCD undergoes a phase 
transition from a confining to a deconfined phase.  Despite the
phenomenological success of QCD, the understanding of the phenomenon of
confinement is still far from being satisfactory. The same is true for the
deconfined phase: the picture of a quark-gluon plasma implies certain
properties of quarks and gluons at high temperatures that have not yet been 
convincingly demonstrated.

This state of affairs is due to the fact that both, confinement and
deconfinement, are not describable by perturbation theory. Thus, investigations
of this and related topics have to rely on non-perturbative methods. Lattice
Monte-Carlo simulations of the Euclidean path integral provide the most direct 
approach and have provided convincing evidence that the pure SU(3) Yang--Mills theory
undergoes a phase transition at a critical temperature $T_c \approx 270$ MeV,
see {\it e.g.\/} \cite{Karsch:2003jg}.
The confining properties as measured by the (temporal) Wilson-loop change 
across this transition which indicates that quarks are, at least partially, 
deconfined in the high-temperature phase. The fate of gluon confinement, on the 
other hand, is much less clear.

At vanishing temperature, gluon confinement has been related to the infrared
behaviour of the gluon and ghost propagators in covariant gauges,  for reviews
see {\it e.g.\/} \cite{Alkofer:2000wg,Roberts:2000aa} or for compact 
presentations of this issue \cite{vonSmekal:2000pz}. These propagators have 
been studied in lattice
calculations, see \cite{Bonnet:2001uh} and references therein,  employing
Renormalisation Group techniques \cite{Pawlowski:2003hq} and Dyson--Schwinger
equations (DSEs), see \cite{Fischer:2003rp} and references therein.
The results of these different methods agree qualitatively and quantitatively:
The gluon propagator is infrared suppressed and the ghost propagator is infrared
enhanced. From the latter property  the Kugo--Ojima confinement
criterion \cite{Kugo:1979gm} follows. Furthermore, it is equivalent to 
Zwanziger's boundary condition on the Gribov horizon \cite{Zwanziger:mf} (see
also \cite{Zwanziger:2003cf}) which guarantees that only
field configurations within the first Gribov horizon \cite{Gribov:1978wm}
contribute. In this approach, the occurrence of gluon confinement is thus
related to the Gribov problem: due to the long-range nature of Faddeev--Popov
ghosts not only Gribov copies are avoided but also the long-range propagation 
of transverse gluons is inhibited. It is therefore interesting to learn
how such a picture changes with temperature.

Here, we will present the solution of coupled DSEs for the gluon and ghost
propagators at non-vanishing temperature $T\neq 0$. We focus on approaching 
the phase transition from the low-temperature confining phase. An investigation 
studying the high-temperature limit has been presented elsewhere \cite{Maas:2004se}.  
A solution on a space-time torus will be given because the use of a 
compact manifold  provides a natural infrared regulator (comparable to the one
present in lattice calculations). 

This paper is organised as follows:
first, we briefly outline the finite-temperature framework for the DSEs.
For the sake of making the presentation self-contained  we discuss in section \ref{sec:truncation}
the truncation scheme employed. 
We then present the numerical results for the propagators. 
These are then used to calculate the effective action
\cite{Luttinger:1960ua,Haeri:hi} in our truncation scheme.  
In the last section we conclude. Technical details are deferred to two
appendices.

\section{Gluon and Ghost Dyson--Schwinger Equations at Non-vanishing
Temperatures}
\label{sec:DSE}

Throughout this investigation the imaginary-time formalism, see {\it e.g.\/} 
\cite{Kap93,Bel96}, is employed. 
This implies using periodic boundary conditions in ``time'' not only for
the gluon fields but also for the ghost fields (despite their Grass\-mann\-ian
nature). 
In four-momentum space this leads to a sum over discrete 
Matsubara frequencies $\omega_n=2\pi n T$ rather than a continuous integral: 
\begin{equation}
\label{measure}
  \int \frac{d^4q}{(2 \pi)^4} f(q) \rightarrow T \sum_{n=-\infty}^{+\infty} 
  \int \frac{d^3\vec q}{(2 \pi)^3} f(\omega_n,\vec q) \; ,
\end{equation}
involving the three-momentum $\vec q$. Here, we have chosen to study the 
system in the rest frame of the heat bath thus working in a formalism 
where Poincar{\'e} covariance is not manifest.   

At non-vanishing temperature the full gluon propagator acquires a more 
complicated tensor structure. In Landau-gauge, which we use throughout this manuscript,
the gluon propagator is purely transverse with respect to the gluon
four-momentum which, at zero temperature, implies that it can be described by one
scalar function. At non-vanishing temperatures two independent tensor
structures exist, one longitudinal and one transverse to the heat bath.
The decomposition for the rest frame of the heat bath \cite{Kap93,Bel96}
is explicitely given by
\begin{eqnarray} 
D^{ab}_{\mu \nu}(k)&=& \frac{\delta^{ab}}{k^2} 
   \bigl(P_{T\, \mu \nu}(k) Z_m(k_0,|\vec k|) \nonumber \\
&& \qquad 
        +P_{L\, \mu \nu}(k) Z_0(k_0,|\vec k|)\bigr) \label{Dmunu} \\
  P_{T\, i j}(k)&=&\delta_{ij} -\frac{k_i k_j}{\vec k^2}, 
  \quad P_{T\, 0 0}=P_{T\, i 0}=P_{T\, 0 i}=0 \quad ,\nonumber \\
  P_{L\, \mu \nu}(k)&=&P_{\mu \nu}(k)-P_{T\, \mu \nu}(k), 
  \quad P_{\mu \nu}=\delta_{\mu \nu}-\frac{k_{\mu} k_{\nu}}{k^2}\; , \label{Projectors} \\ 
  i,j&=&1,2,3; \quad \mu,\nu=1,2,3,4, \nonumber   
\end{eqnarray}
with $k^2=k_0^2+\vec k^2$. The ghost propagator is a Lorentz scalar and does not
acquire further structures:
\begin{equation}
\label{DG} D^{ab}_G(k)=-\frac{\delta^{ab}}{k^2}G(k_0,|\vec k|).
\end{equation}
Another major change in the renormalised DSEs as compared to the ones for zero 
temperature is the substitution (\ref{measure}). The equations for the ghost
and gluon propagator, $D_G(k)$ and $D_{\mu \nu}(k)$ , respectively, read:
\begin{eqnarray}  
D^{-1}_G(k) &=& -\tilde Z_3 k^2 \nonumber \\ 
&+&  g^2 N_C \tilde Z_1  T \sum_{n=-\infty}^{+\infty}
\int \frac{d^3\vec q}{(2 \pi)^3}\nonumber \\ 
&& i k_\mu  D_{\mu \nu}(k-q) G_\nu(k,q)D_G(q) \; ,
\label{eq:ghost}
\end{eqnarray}
\begin{eqnarray} 
D^{-1}_{\mu \nu}(k) &=& Z_3 
\left( \delta_{\mu \nu}-\frac{k_{\mu} k_{\nu}}{k^2} \right) k^2 \nonumber \\ 
&-&g^2 N_C \tilde Z_1 T \sum_{n=-\infty}^{+\infty}
\int \frac{d^3\vec q}{(2 \pi)^3}\nonumber \\ 
&& i q_\mu  D_G(p) D_G(q)G_\nu(p,q) \nonumber \\
&+&\frac12 g^2 N_C Z_1 T \sum_{n=-\infty}^{+\infty} 
\int \frac{d^3 \vec q}{(2 \pi)^3} \nonumber \\ 
&&\Gamma^{(0)}_{\mu \rho \alpha}(k,-p,q) 
 D_{\alpha \beta}(q)D_{\rho \sigma}(p)\Gamma_{\beta \sigma \nu}(-q,p,-k)
\nonumber \\
&+&\ldots \; ,\label{eq:gluon}
\end{eqnarray}
with $p=k+q$. Contributions involving the four-point vertex are not explicitly given here.
The colour structure is also suppressed.
$\Gamma^{(0)}_{\alpha \beta \gamma}$ denotes the tree-level 3-gluon-vertex, 
$\Gamma_{\alpha \beta \gamma}$ the full three-gluon-vertex function
and $G_\nu$ the full ghost-gluon vertex.
Here, the following renormalisation constan\-ts appear: 
$Z_3$ for the gluon wave function, 
$\tilde Z_3 $ for the ghost wave function, 
$Z_1$ for the 3-gluon vertex $\Gamma_{\alpha \beta \gamma}$
and $\tilde Z_1$ for the ghost-gluon vertex $G_\nu$.
In Landau-gauge one has $\tilde Z_1=1$
\cite{Taylor:ff}. This non-renormalisation of the ghost-gluon  
vertex is crucial for the success of the employed truncation scheme
\cite{Watson:2001yv,Lerche:2002ep}. Following \cite{vonSmekal:1997is}
a MOM renormalisation scheme will be used. Thus the renormalisation constants
depend on the renormalisation scale $\mu$ and the ultraviolet cutoff,
for details see below.

At vanishing temperature, the DSEs on a torus provide a good tool to study finite 
volume and periodic boundary conditions effects and allow to compare to results of
continuum methods and the lattice. To extend this approach to finite temperature 
one can directly apply the procedure described in \cite{Fischer:2002eq} by allowing 
for a different size of the torus in time-direction. The corresponding length is 
determined by the temperature, $L_0=\beta=1/T$ $(k_B=1)$ with $L_0 \ll L$, $L$ being 
the size of the torus in the spatial directions. Thus we substitute
\begin{align}
T \sum_{n=-\infty}^{+\infty} \int \frac{d^3 q}{(2 \pi)^3} \rightarrow \frac{T}{L^3} 
\sum_{n=-\infty}^{+\infty} \sum_{j_1,j_2,j_3} \;.
\label{eq:measuretorus}
\end{align}
In order to solve the equations numerically, one needs an ultraviolet momentum 
cutoff $\Lambda$. As we use the same renormalization procedure employed at $T=0$ this 
should be an $O(4)$ invariant cutoff also at finite temperature. 
The integrals are replaced by the Matsubara sums, and the system of equations is solved 
self-consistently for a finite number of sampling points of the dressing functions.
\par
In the limit of infinite spatial volume, the effective temperature on the torus is
\begin{equation}
\label{eq:T} T=1/L_0.
\end{equation}
For small tori this equality can be affected by finite size effects which will be 
discussed in the appendix B, confirming \eqref{eq:T} to be a good estimate.

\section{The truncation scheme}
\label{sec:truncation}
Since the DSEs form an infinite system of hierarchically 
coupled integral equations, one has to truncate the system to make it tractable. 
We employ a truncation scheme which has been tested in various ways at zero 
temperature \cite{Fischer:2002eq}. The consequences of such truncations have been studied extensively in the literature, see ref.\ \cite{Alkofer:2000wg} for a review. In the ultraviolet the truncation is fixed by requiring consistency with  perturbation theory. In the infrared very likely only the behavior of the ghost-gluon vertex is relevant \cite{Watson:2001yv,Lerche:2002ep,Alkofer:2004it}. A bare ghost-gluon vertex as employed here is in agreement with recent studies using DSEs \cite{Schleifenbaum:2004id} and lattice methods \cite{Cucchieri:2004sq}. Thus the truncation is probable even exact in the infrared, as has been argued using stochastic quantization \cite{Zwanziger:2003cf}. Thus only at mid-momenta truncation dependent effects are relevant, but are strongly constrained. The results in the vacuum are furthermore in very good agreement with lattice calculations \cite{Bowman:2004jm}.

In this scheme only the propagators of the ghost 
and the gluon fields are determined self-consistently, while
full vertex functions either have to be constructed or are replaced by bare ones.
As already mentioned, the gauge parameter is fixed to the Landau-gauge value. 
The Landau-gauge is a fixed point of the renormalisation group and thus the gauge 
parameter is not renormalised. Furthermore, 
contributions from the four-gluon vertex are neglected; this is justified for two reasons.
In the first place, the tadpole diagram only contributes a scale-free, divergent constant 
(in Landau-gauge) in the perturbative regime, which drops out by renormalisation. 
Secondly, it can be argued that the two-loop diagrams are sub-leading in the infrared, 
if the 3- and 4-gluon vertices do not acquire highly singular dressing.
Thus we are left with the 3-gluon vertex and the ghost-gluon vertex which are not determined 
by the propagator equations.
Since the ghost-gluon vertex is not ultraviolet divergent in Landau-gauge, replacing the full 
vertex by the tree-level one preserves the one-loop anomalous dimensions of the dressing functions.
In the truncation scheme used, we simply set 
\begin{align} \label{eq:ghostvertex}
G_\nu(q,p) &= i q_\nu .\\
\intertext{For the 3-gluon vertex the following
construction provides the correct one-loop behaviour in the ultraviolet for the propagators without perturbing
their infrared behaviour~\cite{Fischer:2003zc}} 
\label{eq:gluonvertex} \Gamma_{\rho \nu \sigma}(k,p,q)&=F(k,p,q) \Gamma^{(0)}_{\rho \nu \sigma}(k,p,q), \\
F(k,p,q)&=\frac{1}{Z_1(\mu^2,\Lambda^2)} \frac{G(k)^{(-2-6\delta)}}{Z(k)^{(1+3 \delta)}} \times \; \nonumber \\
&~\qquad \qquad \qquad \frac{G(p)^{(-2-6 \delta)}}{Z(p)^{(1+3 \delta)}}\; ,
\end{align}
with
\begin{align} \label{eq:Z}
 Z(k) &:= \frac13 Z_0(k) + \frac23 Z_m(k) \; ,\\ 
 \Gamma^{(0)}_{\rho \nu \sigma}(k,p,q) &=-i(k-p)_\rho \delta_{\mu \nu}-i(p-q)_\mu \delta_{\nu \rho} \nonumber \\
 &-i(q-k)_\nu \delta_{\mu \rho} \; ,
\end{align}
and $\delta=-9/44$ being the anomalous dimension of the ghost propagator.
In the vertex ansatz \eqref{eq:gluonvertex} we have chosen the linear combination 
of eq. \eqref{eq:Z} 
for the two-gluon dressing function, because it corresponds closest to
the zero-temperature dressing of the gluon propagator.
The vertex construction  \eqref{eq:gluonvertex}, taken over from vanishing temperature studies, 
should also be sufficient at finite temperatures, since it mainly  affects the ultraviolet regime
which is nearly temperature independent up to very high temperatures.

The truncation described causes spurious longitudinal contributions to the gluon polarization.
These spurious terms are quadratically divergent.
It is possible to isolate the part without quadratic divergences by contracting with the
traceless Brown-Pennington projector \cite{Brown:1988bm}, 
\begin{equation} \label{eq:brown}
\mathcal{R}(k) = \delta_{\mu \nu} -4\frac{k_\mu k_\nu}{k^2} \; ,
\end{equation}
which projects out terms proportional to $\delta_{\mu \nu}$.
However, this projector interferes with the infrared analysis. 
In order to take care of this problem at zero temperature, the gluon equation is contracted 
with the transverse projector
\begin{equation} \label{eq:transverse}
\mathcal{P}(k) = \delta_{\mu \nu} - \frac{k_\mu k_\nu}{k^2} \; ,
\end{equation}
and a quadratically divergent tadpole-like tensor structure is subtracted from the gluon loop in the gluon DSE.
It is given by \cite{Fischer:2003zc}
\begin{align} \label{tadpole}
  Q_{\mu \nu} = \frac54 \frac{1}{k^2 q^2}  \delta_{\mu \nu} \; .
\end{align}
At non-vanishing temperatures there are other possible quadratically divergent tensor structures.
They mainly gi\-ve different contributions in the infrared which are subleading as long as the
gluon dressing functions are not infrared divergent. The infrared exponents are only weakly dependent on
the actual choice \cite{Fischer:2002eq,Fischer:2003zc}. Thus we subtract the term \eqref{tadpole} to remove all spurious divergences.
We are then left with the intrinsic logarithmic divergences 
which are renormalised in the MOM-scheme as described in ref. \cite{Fischer:2002eq}.

The ans{\"a}tze for the vertices \eqref{eq:ghostvertex} and \eqref{eq:gluonvertex} are inserted into 
eqs. \eqref{eq:gluon} and \eqref{eq:ghost}. 
After contracting the gluon equation with the heat-bath transversal and
longitudinal projectors \eqref{Projectors}, one arrives at the following three coupled equations for
the three scalar dressing functions $G$, $Z_m$ and $Z_0$ defined in eqs. \eqref{Dmunu} and \eqref{DG}:
\begin{align}
\frac{1}{G(k)}&=\tilde Z_3 +g^2N_C\tilde Z_1 T \sum_{n=-\infty}^{+\infty} \int \frac{d^3 q}{(2 \pi)^3}
\frac{G(q)}{k^2 q^2 (k-q)^2} \nonumber \\
&\left\{A_T Z_m(k-q)+A_L Z_0(k-q) \right\} \label{eq:ghostlong} \; , \\ 
\frac{1}{Z_m(k)}&=Z_3 -\frac12 g^2N_C\tilde Z_1 T \sum_{n=-\infty}^{+\infty} \int \frac{d^3 q}{(2 \pi)^3}
\frac{G(q)G(p)}{k^2 q^2 p^2} R   + \nonumber \\
& \frac12 g^2N_C Z_1 T \sum_{n=-\infty}^{+\infty} \int \frac{d^3 q}{(2 \pi)^3} 
\frac{F(q,p,k,Z,G,Z_1)}{k^2 q^2 p^2} \nonumber \\
&\left[M_T Z_m(q)Z_m(p)+ M_1 Z_0(q)Z_m(p)+\right. \nonumber \\  
& \left. M_2 Z_0(p)Z_m(q)+M_L Z_0(q)Z_0(p) \right] \label{eq:gluontransverse} \; ,\\
\frac{1}{Z_0(k)}&=Z_3 -g^2N_C\tilde Z_1 T \sum_{n=-\infty}^{+\infty} \int \frac{d^3 q}{(2 \pi)^3}
\frac{G(q)G(p)}{k^2 q^2 p^2} P  + \nonumber \\
& g^2N_C Z_1 T \sum_{n=-\infty}^{+\infty} \int \frac{d^3 q}{(2 \pi)^3} \frac{F(q,p,k,Z,G,Z_1)}{k^2 q^2 p^2} \nonumber \\
&\left[N_T Z_m(q)Z_m(p)+ N_1 Z_0(q)Z_m(p)+\right. \nonumber \\
&\left.N_2 Z_0(p)Z_m(q)+ N_L Z_0(q)Z_0(p) \right] \label{eq:gluonlongitudinal} \; .
\end{align}
\begin{figure*}
\begin{align*}
\parbox{35mm}
{\epsfig{file=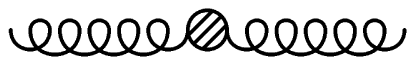,width=35mm}} ^{-1}
&=\; \parbox{35mm}
{\epsfig{file=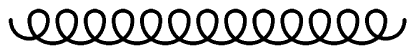,width=35mm}} ^{-1}
\;+\;\parbox{35mm}
{\epsfig{file=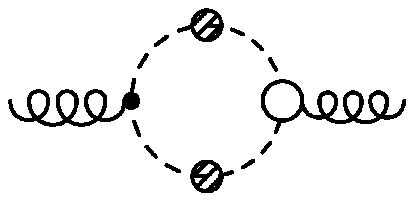,width=35mm}} 
\;-\;\frac12\parbox{35mm}
{\epsfig{file=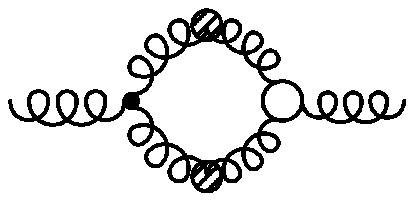,width=35mm}} \nonumber \\
& \; - \frac12 \;
\parbox{35mm}
{\epsfig{file=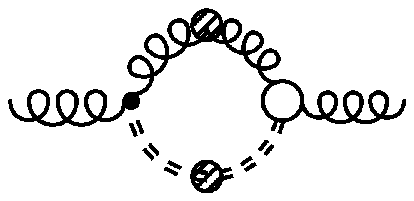,width=35mm}} 
\; - \frac12 \;
\parbox{35mm}
{\epsfig{file=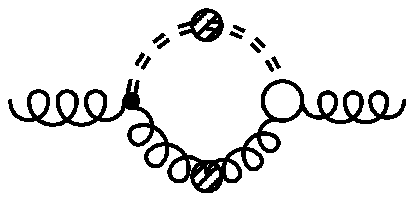,width=35mm}} 
\; - \frac12 \;
\parbox{35mm}
{\epsfig{file=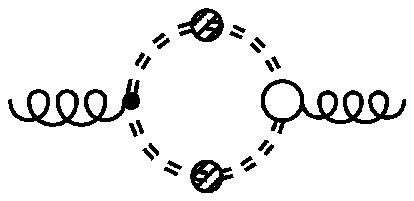,width=35mm}} \nonumber \\
\parbox{35mm}
{\epsfig{file=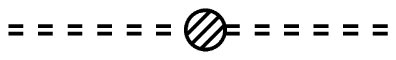,width=35mm}} ^{-1}
&=\;  \parbox{35mm}
{\epsfig{file=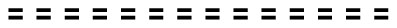,width=35mm}} ^{-1}
\;+\;\parbox{35mm}
{\epsfig{file=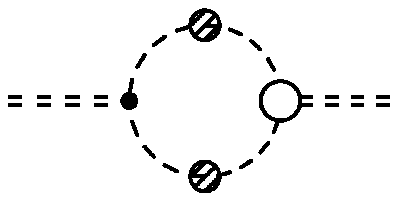,width=35mm}} 
\;-\;\frac12\parbox{35mm}
{\epsfig{file=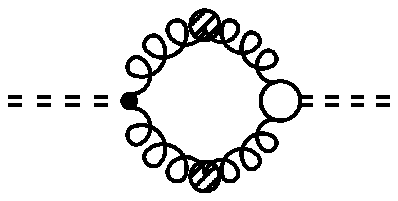,width=35mm}} \nonumber \\
& \; - \frac12 \;
\parbox{35mm}
{\epsfig{file=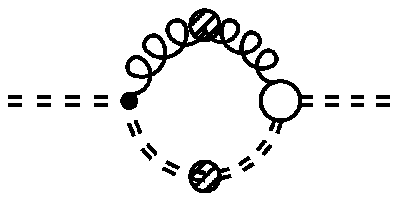,width=35mm}} 
\; - \frac12 \;
\parbox{35mm}
{\epsfig{file=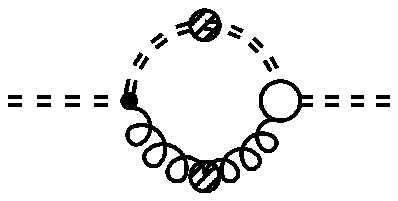,width=35mm}} 
\; - \frac12 \;
\parbox{35mm}
{\epsfig{file=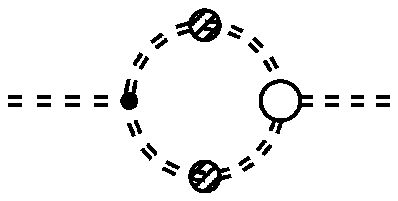,width=35mm}} \nonumber \\
\parbox{35mm}
{\epsfig{file=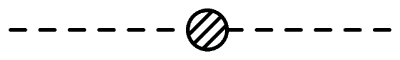,width=35mm}} ^{-1}
&=\;  \parbox{35mm}
{\epsfig{file=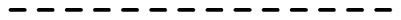,width=35mm}} ^{-1}
\;-\;\parbox{35mm}
{\epsfig{file=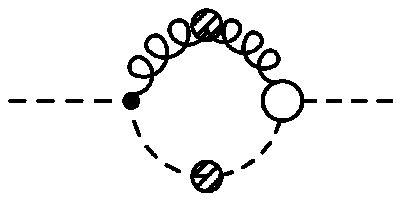,width=35mm}} 
\;-\; \parbox{35mm}
{\epsfig{file=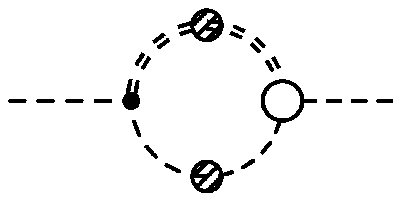,width=35mm}} 
\end{align*}
\caption{Diagrammatic representation of the propagator DSEs in the truncation scheme used here.
Wiggly lines denote heat-bath transverse gluon propagators, double-dashed lines are 
heat-bath longitudinal propagators and dashed lines represent ghost propagators. Blobs
indicate dressed propagators and vertex functions, respectively.}
\label{fig:1}
\end{figure*}
Note that $q_0=\omega_n=2\pi n T$.
Herein colour indices have been contracted and the tensor $\delta^{ab}$ has been separated.
The expressions for the kernel functions 
$A_T$, $A_L$, $R$, $M_T$, $M_1$, $M_2$  $M_L$, $P$ and $N_T$, $N_1$, $N_2$, $N_L$ can be found in appendix A, see  
eqs. \eqref{eq:ghostkernels}-\eqref{eq:gluonkernelslongitudinalend}.
A graphical representation for these equations is given in fig. \ref{fig:1}. 

The MOM-renormalization scheme, applied for vanishing temperature, is used here.
This amounts to solving subtracted equations, {\it i.e.} the respective values at the 
renormalization scale are subtracted on the {\it l.h.s} and {\it r.h.s}. 
The renormalization conditions are then
used to calculate the renormalization constants. 
The introduction of a finite temperature does not give rise to any novel 
divergences \cite{Kap93}. However, finite, temperature-dependent modifications of the renormalization constants 
occur. Thus we impose the same renormalization condition for both the transversal gluon propagator
and the longitudinal one and determine a renormalization constant for each. 

Due to the compactification of space-time, the integrals become discrete, infinite sums. 
To solve the equations numerically one has to introduce a cut-off $\Lambda$ and sum only up to this cut-off. 
The integral equations are solved self-consistently. 
In order to speed up convergence a Newton-Raphson iteration is used. 
As it converges only locally we start with results from
zero temperature and increase the temperature in small steps.
The method will be explained in more detail elsewhere.

\section{Results}\label{sec:res}

\begin{figure}
\begin{center}
  \epsfig{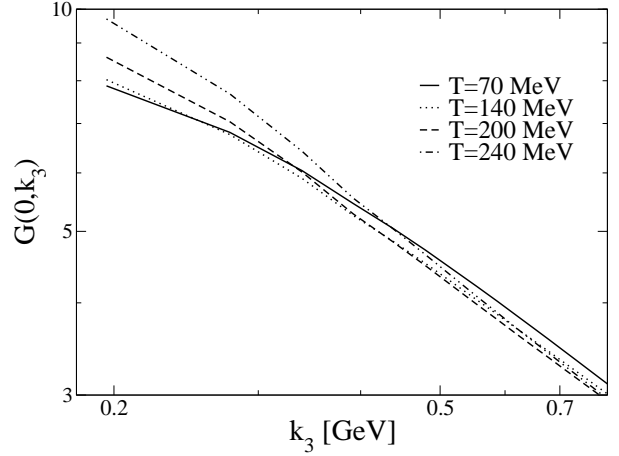}
\end{center}
 \caption{The ghost dressing function $G$ in the infrared region at different temperatures
 ($24^3$ momentum grid).}\label{fig:G_24_70-240}
\end{figure}

\begin{figure}
 \begin{center}
  \epsfig{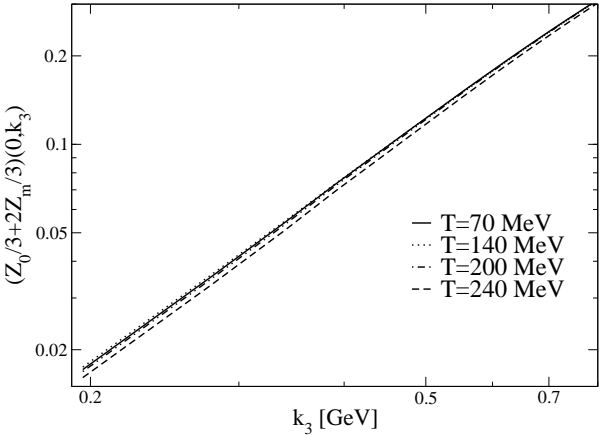}
  \end{center}
  \caption{The averaged gluon dressing function $Z$ in the infrared region at different
  temperatures ($24^3$ momentum grid).}\label{fig:Z_24_70-240}
\end{figure}

\begin{figure}
\begin{center}
  \epsfig{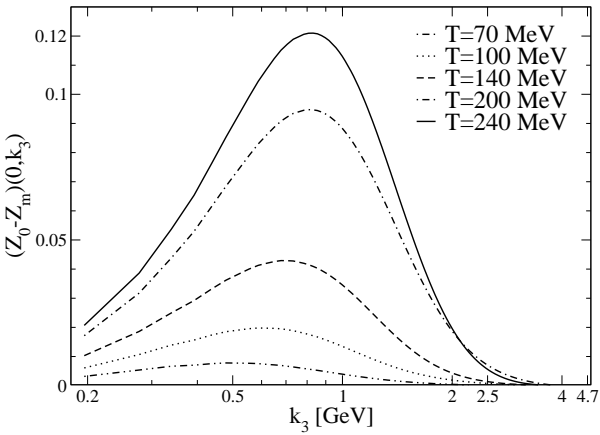}
\end{center}
 \caption{The difference $\Delta Z$ of the gluon dressing functions at different
 temperatures ($24^3$ momentum grid).} \label{fig:DZ_24_70-240}
\end{figure}

In order to display the temperature dependence of the resulting dressing functions 
we will show the linear combination \eqref{eq:Z} of the dressing functions $Z_0$ and $Z_m$ as
well as their difference
\begin{align} \label{eq:DeltaZ}
  \Delta Z &= Z_0 - Z_m. 
\end{align}
In a first step only the zeroth Matsubara mode is discussed here, the other modes will be presented below.
The combination $Z$ in eq. \eqref{eq:Z} is the natural one to compare with zero temperature results, while
$\Delta Z$ measures temperature effects on the tensor structure of the gluon propagator.
The results indicate that $G$ and $Z$ are nearly temperature independent, see figs. \ref{fig:G_24_70-240} and 
\ref{fig:Z_24_70-240}. 
The momentum scale is obtained from the corresponding T=0 calculation \cite{Fischer:2002eq}.
Note that for the $24^3$ grid and the ultraviolet cutoff of $4.7 GeV$ corresponds
to a spatial volume of approximately 6 fm$^3$.
Unfortunately, it is not possible to extract reliable infrared exponents from these torus results.
Hence, the chan\-ges can be either due to changes in the infrared coefficient or in the exponent.
The ghost dressing would suggest a change in the exponent, while the gluon dressing points to
a change in the coefficient. Nevertheless, the results allow to conclude, that the power laws persist in the 
infrared regime. The changes of the ghost dressing functions with respect to temperature 
are more likely fluctuations due to finite-size effects rather than an intrinsic temperature effect.
Fig. \ref{fig:DZ_24_70-240} displays, that $\Delta Z$  is temperature dependent to a rather large extent 
at intermediate momenta. The temperature dependences of $Z_0$ and
$Z_m$ have opposite signs and $\Delta Z$ is of the order of a few percent in the intermediate momentum 
region compared to the sum. Up to about $1$ GeV the shape of $\Delta Z$ can be explained
by just changing the infrared coefficient and then connecting continuously to the 
ultraviolet regime, where $\Delta Z$ has to vanish. 
The data presented here are for a cutoff of $4.7$ GeV. For other cutoffs the results vary slightly,
in a range which is expected for such low cutoffs.


\begin{figure}
\begin{center}
  \epsfig{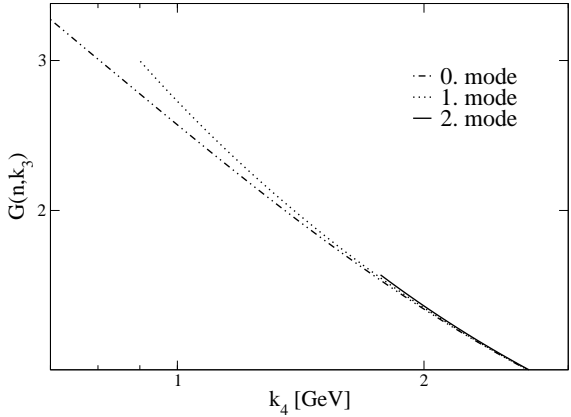}
\end{center}
 \caption{Different Matsubara modes of the ghost dressing function at $T=140$ MeV 
 ($24^3$ momentum grid).}\label{fig:G_140_modes}
\end{figure}

\begin{figure}
 \begin{center}
  \epsfig{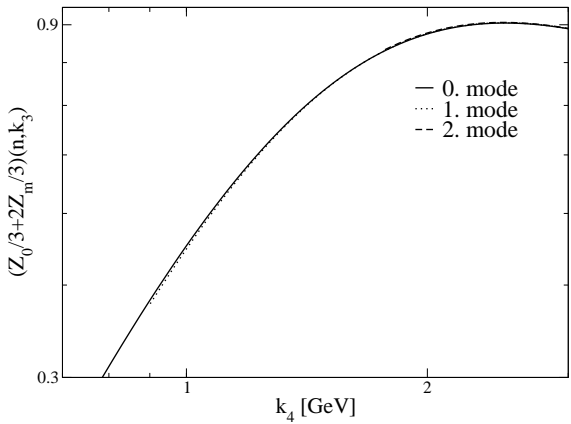}
  \end{center}
  \caption{Different Matsubara modes of the averaged gluon dressing function $Z$ at 
  $T=140$ MeV ($24^3$ momentum grid).}\label{fig:Z_140_modes}
\end{figure}

\begin{figure}
\begin{center}
  \epsfig{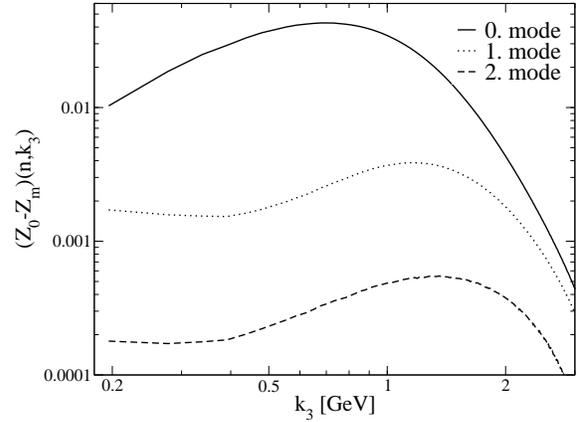}
\end{center}
 \caption{Different Matsubara modes of the difference $\Delta Z$ of the gluon 
 dressing functions at $T=140$ MeV ($24^3$ momentum
 grid).} \label{fig:ZM_140_modes}
\end{figure}

For the higher Matsubara modes the dressing functions have an interesting
property:
the combined gluon dressing $Z$ and the ghost dressing $G$ seem to depend only on
the 4-momentum (at least approximately), while the higher modes of $\Delta Z$ 
are not $O(4)$ invariant, as expected  (see figs.~\ref{fig:G_140_modes}-\ref{fig:ZM_140_modes})
\footnote{The first Matsubara mode of the ghost dressing (see fig. \ref{fig:G_140_modes})
deviates slightly from $O(4)$ invariance. This is due to the torus regularization.}.

\begin{figure}
\begin{center}
  \epsfig{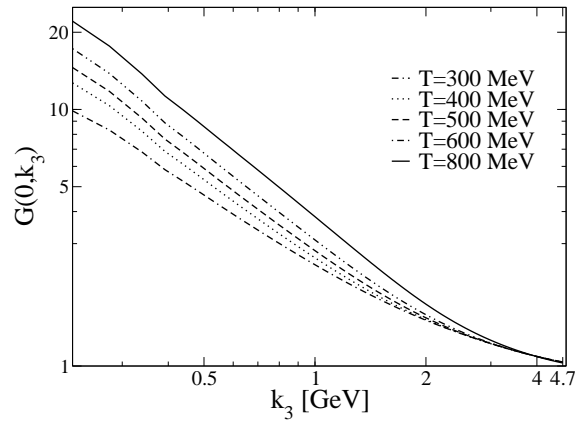}
\end{center}
 \caption{The ghost dressing function $G$ at high temperatures ($24^3$
 momentum grid).} \label{fig:G24_300-800}
\end{figure}

\begin{figure}
\begin{center}
  \epsfig{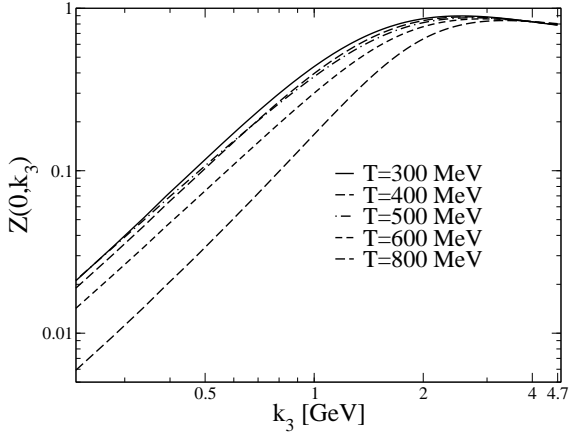}
\end{center}
 \caption{The averaged gluon dressing function $Z$ at high temperatures ($24^3$
 momentum grid).} \label{fig:Z24_300-800}
\end{figure}

Due to the numerical method, we are biased to stay in the confining phase. Nevertheless 
it is surprising that we find qualitatively similar solutions also for high temperatures. 
These solutions are numerically stable up to temperatures allowed by the 
UV cut-off, $2\pi T \approx\Lambda$. In order to draw conclusions on the character of the phase transition we need
results from other numerical methods to support this finding.  If so, our results would indicate a first order phase
transition.
Figs. \ref{fig:G24_300-800} and \ref{fig:Z24_300-800} show that the changes of the "wrong-phase" solutions
with temperature are also smooth and are qualitatively the same as for small temperatures. 

The running coupling $\alpha(k^2)$, being a renormalization group invariant quantity, is of special interest.
In Landau-gauge at zero temperature, a non-perturbative running coupling $\alpha(p^2)$ can be defined via the
relation:
\begin{align}
 \alpha(p^2)=\alpha(\mu^2)G^2(p^2,\mu^2)Z(p^2,\mu^2) \; .
\end{align}
In order to smoothly connect to the results for vanishing temperature we take for $Z$
the linear combination \eqref{eq:Z}:
\begin{align} \label{eq:alphaT}
 \alpha(n,p_3^2,T):=\alpha(\mu^2)G^2(n,p_3^2,\mu^2,T)Z(n,p_3^2,\mu^2,T).
\end{align}
As expected from the results for the dressing functions, the coupling \eqref{eq:alphaT} shows no
temperature dependence beyond numerical uncertainties.



%\newpage
\section{CJT Action and Phase Transition} \label{sec:cjt}
\begin{figure*}
\begin{align}
V_2(D_{\mu \nu},D_G)&=
\; \frac12 \;
\parbox{35mm}
{\epsfig{file=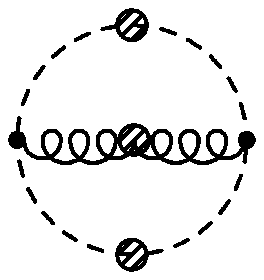,width=35mm}}
\;-  \frac{1}{12} \;
\parbox{35mm}
{\epsfig{file=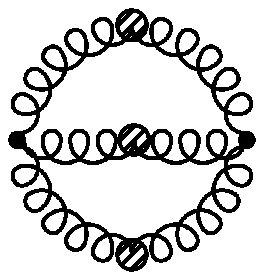,width=35mm}}
\nonumber
\end{align}
\caption{Diagrammatic representation of the two-loop contributions to the effective action.}
\label{fig:cjtinteracting}
\end{figure*}

\begin{figure*}
\begin{align}
V_2(D_{\mu \nu},D_G)&=
\frac13
\left(
\parbox{35mm}
{\epsfig{file=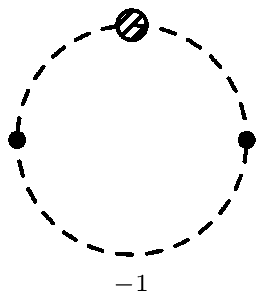,width=35mm}}
-1 \right)
\; - \frac16 \;
\left(
\parbox{35mm}
{\epsfig{file=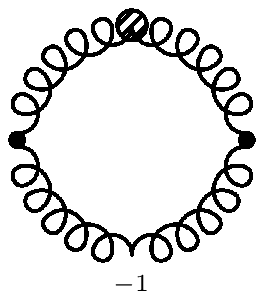,width=35mm}}
-1 \right)  \nonumber
\end{align}
\caption{$V_2(D_{\mu \nu},D_G)$ after inserting the reintegrated DSEs.}
\label{fig:13}
\end{figure*}
As we want to investigate the phase transition it is necessary to compute  
thermodynamic quantities.
In order to do so, we need the temperature dependent effective action for
vanishing fields and full propagators. 
Since we want to calculate the temperature dependence of different quantities, we need
to assure energy conservation.
Thus the DSEs, which we use to compute the dressing functions, have to be the variational 
equations of the energy functional, i.e. of the effective action.
We take the known 2PI effective action by Cornwall, Jackiw and Tomboulis (CJT) \cite{Luttinger:1960ua}
for fully dressed propagators and bare vertices
\footnote{Recently there has been renewed interest in the formalism also for higher particle irreducibilities \cite{Berges:2004pu},
however, the corresponding treatment is beyond the scope of this paper. Anyways, our truncation scheme
only includes 1-loop graphs for the propagator DSEs, so the action contains at most two-loop diagrams and
is thus 2PI.}.
First we discuss the action for vanishing temperature, given by 
the formula (c.f. ref.~\cite{Haeri:hi})
\begin{align}
\label{cjtfull}
V(D_{\mu \nu},D_G)&= V_0(D_{\mu \nu},D_G)+V_2(D_{\mu \nu},D_G) \; , \\
V_0(D_{\mu \nu},D_G)&= \int \frac{d^4p}{(2\pi)^4} {\rm Tr}
\left\{ \frac12 \left[{D^0_{\mu \alpha}}^{-1}(p)D_{\alpha \nu}(p) -\right. \right. \nonumber \\ 
&\left. g_{\mu \nu} \right] -\frac12 \ln \left({D^0_{\mu \alpha}}^{-1}(p)D_{\alpha \nu}(p) \right)- \nonumber \\
& \left[D_G^0(p)^{-1}D_G(p)-1 \right] \nonumber \\
&+\ln \left(D_G^0(p)^{-1}D_G(p) \right) 
\Big\} \; ,
\end{align}
where we already employed Euclidean space-time conventions.

In a truncation scheme with bare vertex functions only the diagrams of fig. \ref{fig:cjtinteracting} contribute to 
the effective action. For our truncation to the gluon propagator DSE, the 3-gluon vertex \eqref{eq:gluonvertex} is 
constructed  from the dressing functions of the propagators. However, throughout this investigation we neglect variations of the 
vertex with respect to the gluon and ghost dressing functions, and thus employ 
\begin{align}
 V_2(D_{\mu \nu},D_G)&= \frac12 \int \frac{d^4p}{(2\pi)^4}  \int \frac{d^4q}{(2\pi)^4}     
 i p_\mu  D_{\mu \nu}(p-q) \nonumber \\
 & i q_\mu D_G(q)D_G(p) - \frac{1}{12} \int \frac{d^4p}{(2\pi)^4}   \int
 \frac{d^4q}{(2\pi)^4} \Big\{ \nonumber \\
& \Gamma^{0}_{\mu \rho \alpha}(p,-p-q,q) D_{\alpha \beta}(q)D_{\rho \sigma}(p+q) \nonumber \\
&\Gamma^{0}_{\beta \sigma \nu}(-q,p+q,-p)D_{\mu \nu}(p)
\Big\} \; .
\end{align}
The variation of the action \eqref{cjtfull} with respect to the propagators reproduces the Dyson-Schwinger equations
for the gluon and ghost propagators \eqref{eq:gluon} and \eqref{eq:ghost}, however, with bare 3-gluon vertex. 
Multiplying the gluon DSE with the full gluon propagator and the ghost DSE with the full ghost propagator 
and performing the trace in colour and Lorentz indices and the integral in momentum space, we recover
the two-loop diagrams of the CJT action.
This results in: 
\begin{align}
V_2(D_{\mu \nu},D_G)&=
\int \frac{d^4p}{(2\pi)^4} {\rm Tr}
\left\{ -\frac16 \left[{D^0_{\mu \alpha}}^{-1}(p)D_{\alpha \nu}(p) -\right. \right. \nonumber \\ 
&\left. \delta_{\mu \nu} \right] + \frac13 \left[D_G^0(p)^{-1}D_G(p)-1 \right] \Big\} \; ,
\label{V2cjtstat}
\end{align}
see fig. \ref{fig:13} for a graphical representation.
Inserting \eqref{V2cjtstat} and the explicit expression for the propagators in dependence on the dressing
functions in \eqref{cjtfull} one obtains:
\begin{align}
V(Z,G)&= N_C\int \frac{d^4p}(2\pi)^4 \Big\{ \nonumber \\ 
& \left(Z_3 Z(p^2)-1 \right) -\frac32 \ln\left(Z_3 Z(p^2)\right) \nonumber \\ 
&  - \frac23\left(\tilde Z_3 G(p^2)-1 \right)+\ln\left(\tilde Z_3 G(p^2)\right)
\Big\}. \label{eq:cjtdyn}
\end{align}
For solving the corresponding DSEs numerically, we introduce a momentum cut-off $\Lambda$. Thus
the renormalization constants $Z_3$ for the gluon wave function and $\tilde Z_3$ for the ghost
propagator not only depend on the renormalization scale $\mu$ but also on the cut-off
\footnote{The propagators are cut-off independent within numerical accuracy as soon as $\Lambda>15$ GeV.}.
Defining $x:=p^2$ one has:
\begin{align}
V(Z,G)&= \frac{N_C}{(4 \pi)^2}\int_0^{\Lambda^2} dx x 
\Big\{ \left(Z_3(\mu,\Lambda) Z(x,\mu)-1 \right)  \nonumber \\ 
& -\frac32 \ln\left(Z_3(\mu,\Lambda) Z(x,\mu)\right) \nonumber \\ 
&  - \frac23\left(\tilde Z_3(\mu, \Lambda) G(x,\mu)-1 \right) \nonumber \\
& +\ln\left(\tilde Z_3(\mu,\Lambda) G(x,\mu)\right) \Big\}.
\end{align}
This quantity diverges
like $\Lambda^4$, if the renormalization constants are only known to a finite accuracy.
Furthermore, the three-gluon vertex is dressed by construction, which is not accounted for by the above action.
In fact the following fit 
\begin{align}
 f(x)&=ax^2+c 
\intertext{to the cut-off dependence of the action yields:}
 a&= -3.2 \times 10^{-3} \nonumber \\
 %b&=2
 c&= (9.8 \pm 1.3) \times 10^{-10}  \textrm{GeV}^4 
\end{align}
The result should be interpreted as vanishing of the effective action as the first term is of purely numerical
origin and the second term is much too small to correspond to a scale of the system, $c^{\frac14} \approx 5.6$ MeV.

At finite temperatures, the integral measure in eq. \eqref{eq:cjtdyn} has to be replaced by
that for the space-time torus, see eq. \eqref{eq:measuretorus}.
Furthermore, the finite temperature gluon propagator with its additional tensor structure \eqref{Dmunu}
has to be inserted.
Finally we get for the temperature dependent 1-particle irreducible CJT-Action:
\begin{align}
V(T)&= N_C T \sum_{n=-\infty}^{+\infty}   \int \frac{d^3\vec q}{(2 \pi)^3}
\left\{ \right. \nonumber \\ 
& \frac23 \left(Z_{3 m} Z_m(\omega_n,|\vec p|)-1 \right) - \ln\left(Z_{3
m} Z_m(\omega_n,|\vec p|)\right) +\nonumber \\ 
& \frac13 \left(Z_{3 0} Z_0(\omega_n,|\vec p|)-1 \right) - \frac12 \ln\left(Z_{3
0} Z_0(\omega_n,|\vec p|)\right) - \nonumber \\ 
& \frac23\left(\tilde Z_3 G(\omega_n, |\vec p|)-1 \right)+\ln\left(\tilde Z_3 G(\omega_n,|\vec p|)\right)
\Big\}\; .
\end{align}
This effective action is temperature independent within numerical uncertainties up to $T=250$ MeV. For higher temperatures large
variations are seen. Whether this is connected to the phase transition remains to be seen.

\section{Conclusions}
In this paper we have extended a truncation scheme for the Dyson-Schwinger equations of
Yang-Mills theories in Landau-gauge to non-vanishing temperatures using the imaginary time
formalism for quantum field theories in thermal equilibrium.
We have obtained numerical solutions for the gluon and ghost propagators.
Furthermore we have derived an expression for the CJT action of the interacting Yang-Mills theory, depending on
dressed propagators.

The results can be summarised as follows:
Temperature dependences of the ghost and gluon propagators are rather weak and the power-law 
behaviour in the infrared, already observed at vanishing temperature, persists. 
The corresponding CJT action vanishes (within numerical uncertainties) and does not possess any
significant temperature dependence up to $T=250$ MeV.
Thus gluons and ghost neither build up any pressure nor give raise to finite entropy of the system.
An improved numerical method, especially one suitable for infinite volume, is, of course, desirable to 
study thermodynamic quantities below the phase transition. Corresponding investigations are
currently performed.

\section*{Acknowledgments}
We thank C.~S.~Fischer, H.~Reinhardt, S.~M. Schmidt, and P.~Maris for fruitful discussions.
We are grateful to C.~S.~Fischer and J.~M.~Pawlowski for a critical reading of the manuscript.
This work has been supported by the DFG under contract GRK683 (European Graduate School T{\"u}bingen-Basel),
by the BMBF (grant number 06DA917) and by the Helmholtz association (Virtual Theory Institute VH-VI-041).

\section*{Appendix}

\section*{A Integral kernels for the DSEs}
\label{sec:appIntegralKernels}

The complete algebraic expressions for the Lorentz structures that occur in the 
DSEs for the different
products of dressing functions, are given here.
First, there are the two kernels of the ghost equation \eqref{eq:ghostlong}:
\begin{align} \label{eq:ghostkernels}
A_T(k,q)& = 
-\frac{\vec k^2 \vec q^2 - (\vec k \cdot \vec q)^2}{(\Vec k-\Vec q)^2} \\
A_L(k,q)&=  -\frac{k^2  q^2 - (k q)^2}{( k- q)^2} +
\frac{\Vec k^2 \Vec q^2 - (\Vec k \cdot \Vec q)^2}{(\Vec k-\Vec q)^2} .
\end{align}
Secondly, the kernel of the ghost loop and the four kernels of the gluon loop for the 
heat-bath transversal part of the gluon equation \eqref{eq:gluontransverse}:
\begin{align} \label{eq:gluonkernelstransverse}
R(k,q)&=-\frac{\left(\vec q^2 \vec k^2 -(\vec k \cdot \vec q)^2 \right)}{\vec k^2} \\
M_T(k,q)&=-2\frac{\vec q^2 \vec k ^2 -(\vec k \cdot \vec q)^2}{\vec k^2 \vec q^2 \vec p^2} 
\left((\vec k \cdot \vec q)^2+\vec k^2 \vec q^2 + \right. \nonumber \\ 
&\left. 2\vec p^2(\vec k^2 + \vec q^2 )\right) \\
M_1(k,q)&=-2 \frac
{  \left(q_0 \vec k \cdot \vec q - k_0 \vec q^2 \right)^2
\left((\vec k \cdot \vec p )^2 + \vec k^2 \vec p^2\right) }
{\vec k^2 \vec q^2  \vec p^2 q^2}\\
M_2(k,q)&=-2 \frac
{  \left(p_0 \vec k \cdot \vec p - k_0 \vec p^2 \right)^2
\left((\vec k \cdot \vec q )^2 + \vec k^2 \vec q^2\right) }
{\vec k^2 \vec q^2  \vec p^2 p^2}\\
M_L(k,q)&= -2\frac{\left(\vec q^2 \vec k^2-(\vec k \cdot \vec q)^2 \right)
\left(q^2 p^2-q_0 p_0 qp \right)^2}
{\vec k^2 \vec q^2 \vec p^2 k^2 p^2} .
\end{align}
And third the corresponding kernels for the heat-bath longitudinal part \eqref{eq:gluonlongitudinal}:
\begin{align} \label{eq:gluonkernelslongitudinal}
P(k,q)&=-\frac{\left(q_0\vec k^2  -k_0\vec k \cdot \vec q \right)^2}{k^2 \vec k^2} \\
N_T(k,q)&=-2 \frac
{  \left(k_0 \vec k \cdot \vec q - q_0 \vec k^2 \right)^2
\left((\vec q \cdot \vec p )^2 + \vec q^2 \vec p^2\right) }
{\vec k^2 \vec q^2 \vec p^2 k^2}\\
N_1(k,q)&= -2\frac{ \left( \vec q^2 \vec k^2 - (\vec k \cdot \vec q)^2  \right)
\left(k^2 q^2-k_0 q_0 kq \right)^2}
{\vec k^2 \vec q^2 \vec p^2 k^2 p^2} \\
N_2(k,q)&= -2\frac{ \left(\vec q^2 \vec k^2 -(\vec k \cdot \vec q)^2  \right)
\left(k^2 p^2-k_0 p_0 kp \right)^2}
{ \vec k^2 \vec q^2 \vec p^2 k^2 q^2} \\
N_L(k,q)&=-\frac{1}{2 \vec k^2 \vec q^2 \vec p^2 k^2 q^2 p^2}
\left[(\vec p \cdot \vec k)\vec q^2\left(k_0p^2+p_0k^2\right)-\right. \nonumber \\ 
&\left. (\vec p \cdot \vec q)\vec k^2\left(p_0q^2+q_0p^2\right)+ \right. \nonumber \\
&\left. (\vec k \cdot \vec q)\vec p^2\left(k_0q^2-q_0k^2\right) \right]^2.\label{eq:gluonkernelslongitudinalend}
\end{align}

\section*{B Finite size effects} \label{sec:appFiniteSize}

\begin{figure}
 \begin{center}
  \epsfig{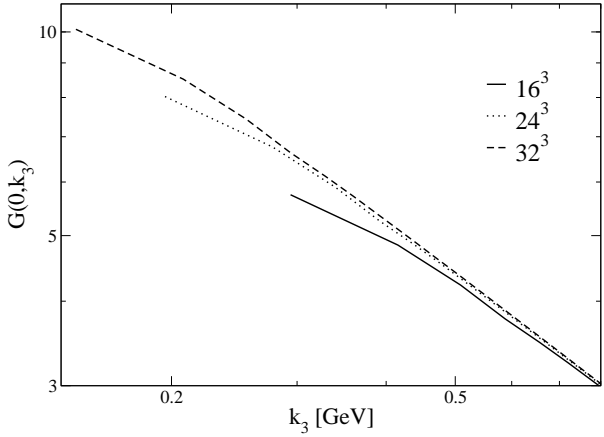}
  \end{center}
  \caption{The ghost dressing function $G$ at $T=140$ MeV on different momentum grids.}\label{fig:G_140_fs}   
\end{figure}

\begin{figure}
 \begin{center}
  \epsfig{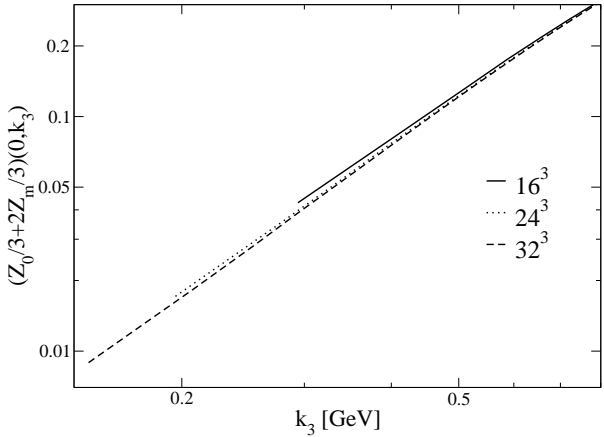}
  \end{center}
  \caption{The averaged gluon dressing function $Z$ at $T=140$ MeV on different momentum grids.}\label{fig:Z_140_fs}
\end{figure}

\begin{figure}
 \begin{center}
  \epsfig{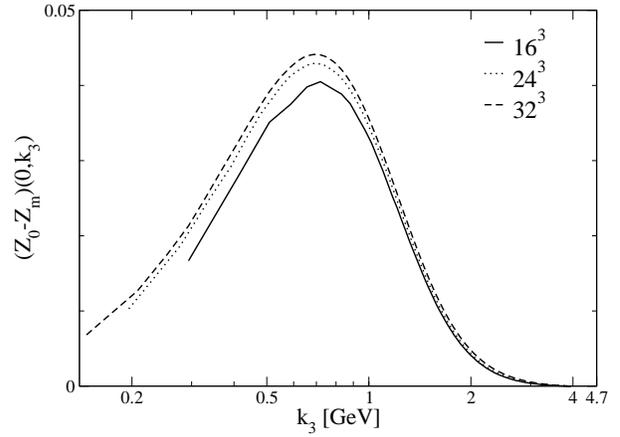}
  \end{center}
  \caption{The difference $\Delta Z$ of the gluon dressing functions at $T=140$ MeV on different momentum grids.}\label{fig:ZM_140_fs}
\end{figure}

As expected, finite-size effects cannot be neglected completely but are rather small and
change the results only in a quantitative way, as can be seen from the figures \ref{fig:G_140_fs} -
\ref{fig:ZM_140_fs}.
The deviations tend to increase for higher temperatures, since the ultraviolet cutoff is rather
low ($4.7$ GeV).
Finite size effects do not alter our conclusions. Especially, the temperature estimate of eq. \eqref{eq:T} is
fairly independent of the finite-momentum lattice size.

\end{document}